\documentclass[sigconf]{acmart}

\AtBeginDocument{%
  \providecommand\BibTeX{{%
    \normalfont B\kern-0.5em{\scshape i\kern-0.25em b}\kern-0.8em\TeX}}}

\copyrightyear{2023}
\acmYear{2023}
\setcopyright{rightsretained}
\acmConference[WWW '23]{Proceedings of the ACM Web Conference 2023}{May 1--5, 2023}{Austin, TX, USA}
\acmBooktitle{Proceedings of the ACM Web Conference 2023 (WWW '23), May 1--5, 2023, Austin, TX, USA}
\acmDOI{10.1145/3543507.3583431}
\acmISBN{978-1-4503-9416-1/23/04}

\usepackage{xcolor}
\usepackage{multirow}
\usepackage{subfig}
\usepackage{bm}
\usepackage{color}
\usepackage{pifont}

\begin{document}

\title[Multiview Representation Learning from Crowdsourced Triplet Comparisons]{Multiview Representation Learning from \\ Crowdsourced Triplet Comparisons}

\author{Xiaotian Lu}
\affiliation{%
  \institution{Kyoto University}
  \city{Kyoto}
  \country{Japan}}
  \email{lu@ml.ist.i.kyoto-u.ac.jp}
  \orcid{0000-0002-4118-2301}
  \authornote{Corresponding author}

  \author{Jiyi Li}
\affiliation{%
  \institution{University of Yamanashi}
  \city{Kofu}
  \country{Japan}}
  \email{jyli@yamanashi.ac.jp}
  \orcid{0000-0003-4997-3850}

 \author{Koh Takeuchi}
\affiliation{%
\institution{Kyoto University}
  \city{Kyoto}
  \country{Japan}}
  \email{takeuchi@i.kyoto-u.ac.jp}
 \orcid{0000-0002-3245-888X}
 
 \author{Hisashi Kashima}
\affiliation{%
\institution{Kyoto University}
  \city{Kyoto}
  \country{Japan}}
  \email{kashima@i.kyoto-u.ac.jp}
 \orcid{0000-0002-2770-0184}

\begin{abstract}
Crowdsourcing has been used to collect data at scale in numerous fields. Triplet similarity comparison is a type of crowdsourcing task, in which crowd workers are asked the question ``among three given objects, which two are more similar?'', which is relatively easy for humans to answer. However, the comparison can be sometimes based on multiple views, i.e., different independent attributes such as color and shape. Each view may lead to different results for the same three objects. Although an algorithm was proposed in prior work to produce multiview embeddings, it involves at least two problems: (1) the existing algorithm cannot independently predict multiview embeddings for a new sample, and (2) different people may prefer different views. 
In this study, we propose an end-to-end inductive deep learning framework to solve the multiview representation learning problem. The results show that our proposed method can obtain multiview embeddings of any object, in which each view corresponds to an independent attribute of the object. We collected two datasets from a crowdsourcing platform to experimentally investigate the performance of our proposed approach compared to conventional baseline methods. 
\end{abstract}



\maketitle
\section{Introduction}
In recent years, deep learning methods have been widely adopted in various fields, and have exhibited remarkable performance~\cite{socher2013recursive,szegedy2015going}. However, these applications largely depend on the collection of sufficient amounts of appropriate training data. Crowdsourcing is an efficient and economical approach in which various data are collected by applying human intelligence~\cite{lu2021crowdsourcing}; asking crowd workers to annotate labels for specified objects in digital images by choosing one of several categories is among the most popular tasks~\cite{kovashka2016crowdsourcing}. However, collecting labels by making choices can become difficult in some cases; the set of all possible categories may not be available in advance, objects may be difficult to recognize, and crowd workers may not be able to provide accurate answers to some questions without expert or professional knowledge.

\begin{figure}[!t]
\centering
\centerline{\includegraphics[width=3.2in]{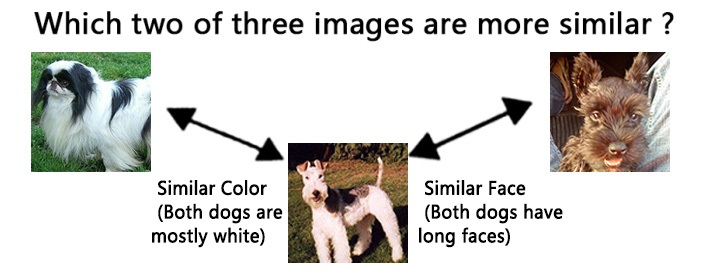}}
\caption{An example of triplet similarity comparison with three samples. Crowd workers are asked to select two similar images among the given three images. In this example, there are two possible views: color and face.}
\Description{An example of triplet similarity comparison with three samples.}
\label{multi-comp}
\end{figure}

To solve this problem, we consider tasks that ask about similarity rather than requesting workers to perform categorization. For example, it is difficult for people with only general knowledge to identify all dog breeds so that the images cannot be directly labeled but similarity data can be collected by comparisons as shown in Figure~\ref{multi-comp}. Similarity comparison data can be used to train representation learning models and provide multi-dimensional vector embeddings of objects. Usually, we expect the similarity (inversely proportional to distance) of embeddings of similar objects to be larger and vice-versa.

Pairwise and triplet similarity comparisons are two representative types of similarity comparison tasks~\cite{heikinheimo2013crowd,tamuz2011adaptively,rankingclustering,li2021label}. Pairwise similarity comparison, also known as absolute similarity comparison, asks crowd workers to answer the following question, i.e., \textbf{\textit{"Are objects A and B similar?"}}.

However, applying pairwise similarity comparisons in crowdsourcing involves at least one key issue; that is, making absolute decisions is generally challenging for humans. This problem is especially noticeable for making subjective judgments. Different crowd workers may have different thresholds of similarity or dissimilarity, thus their judgments conflict with each other often.

Triplet similarity comparisons can be used to solve this problem with pairwise similarity comparison in crowdsourcing, also known as relative similarity comparison. Triplet similarity comparison involves the selection of two relatively similar objects among three that are provided. Here, crowd workers are asked to answer the following question, i.e., \textbf{\textit{"Which two objects among A, B, and C are more similar?"}}.

There are three possible answers: (1) A and B are more similar, (2) A and C are more similar and, (3) B and C are more similar. Triplet similarity comparisons are more accurate than pairwise similarity comparisons because relative comparisons are often easier for humans for common cases~\cite{li2021label}.

However, triplet similarity comparisons can be more ambiguous in some cases, i.e., an object may have multiple attributes. Humans would thus naturally compare them differently in terms of different attributes. In this study, we refer to these characteristics as "views" for simplicity. At least two problems arise in considering multiple views: \\
(1)
First, \textit{different crowd workers consider different views for the same task.} 
For example, the triplet similarity comparison task shown in Figure~\ref{multi-comp}  involves two possible views: color of dogs and face shape of dogs. The leftmost dog and the rightmost dog are clearly dissimilar.
Crowd workers who focus more on color will consider that the leftmost and the middle dogs are more similar.
However, crowd workers who focus more on face shape will consider that the rightmost and the middle dog are more similar. Neither choice is wrong, and we describe this difference as being caused by the different views of the workers in considering different attributes of objects. \\
(2) 
Second, \textit{the same crowd worker might choose different views in different tasks or a different times.} 
A worker might select different views in different situations, e.g., sometimes focusing on color and considering shape in other instances. Moreover, workers may select a view that simplifies their decision making process for a given task.

In this study, we propose a novel end-to-end representation learning framework to learn from multiview triplet data by adding multiple branches to the existing network structure. Our proposed method allows different workers to choose different views for different tasks. We recruited crowd workers and conducted simulation experiments to evaluate the performance of our proposed method.

The contributions of this study are summarized as follows. 

\begin{itemize}
\item In contrast to previous work~\cite{amid2015multiview}, our proposed method performs inductive learning and can provide multiview embeddings for an arbitrary new sample. 

\item To address the problem of different crowd workers responding with different views, we added worker models to reflect the preference of workers for different views.   

\item To address the problem that a given crowd worker might choose different views, the proposed approach adopts triplet entropy to measure the difficulty of deciding on a view. 

\item We used multiple evaluation metrics to evaluate the performance of our proposed approach compared to some existing baseline methods. Additionally, we also confirmed the semantic meaning of multiview embeddings using visualization techniques. 
\end{itemize}

\begin{table*}[!tbp]
 \setlength{\tabcolsep}{2pt}

\caption{Differences between the present work and existing methods. This table indicates the novelty and contribution of this work. "Multiple crowd workers" refers to whether crowdsourcing is involved and whether the difference between workers is considered. "Multiview" refers to whether data are processed in multiple views. "Annotation type" refers to the ground-truth data in supervised learning. "Output" refers to the final desired result. "Inference type“ refers to the mode of learning, whereas inductive learning 
refers to general learning tasks, in which general patterns are learned from training data and applied to testing data, and transductive learning refers to reasoning from training data to specific testing data.}

\begin{center}

\begin{tabular}{ c c c c c c }
\toprule
Study                 & Multiple crowd workers & Multiview & Annotation type       & Output                 & Inference type \\ \midrule
\textbf{Ours}        & \textbf{\ding{51}}          & \textbf{\ding{51}}   &\textbf{Triplet similarity} & \textbf{Multiview Embeddings} & \textbf{Inductive} \\ \midrule
MVTE~\cite{amid2015multiview}    & \ding{55}                  & \ding{51}            & Triplet similarity    & Multiview Embeddings   & Transductive   \\ \hline
Triplet Network~\cite{hoffer2015deep}       & \ding{55}                  & \ding{55}             & Class label           & Embeddings             & Inductive      \\ \hline
Crowd Layer~\cite{rodrigues2018deep,SpeeLFC,conal,li2022beyond}            & \ding{51}                    & \ding{55}             & Class label           & Class Labels           & Inductive      \\ \hline
Multiview Clustering~\cite{zhang2021multi,yao2019multi,zhou2017randomized} & \ding{55}                     & \ding{51}            & N/A                   & Clusters               & Transductive   \\ \hline
Multi-label Learning~\cite{liu2021emerging,sun2010multi,ridnik2021asymmetric}  & \ding{55}                     & \ding{55}             & Multiple class labels & Multiple class labels & Inductive      \\ \bottomrule
\end{tabular}

\label{tab1}

\end{center}
\end{table*}

\section{Related Work}
We briefly review some relevant studies and describe their differences from our proposed method, which are summarized in Table~\ref{tab1}.
\subsubsection*{Multiview Triplet Embedding: }
Amid et al.~\cite{amid2015multiview} proposed an algorithm that produces multiview triplet embeddings (MVTE), in which each view corresponds to a hidden attributes. The input of this algorithm is only the triplet comparison data, which does not contain information about the original object, such as raw images, and the output is the multiview embeddings of each sample. 

The problem can be summarized as follows. 

Given a set of triplets $\mathcal{T}$, each element in $\mathcal{T}$ corresponds to a triplet ordered tuple $(i,j,k)$ in which object $i$ is more similar to object $j$ than object $k$, find $V$-views embeddings of $N$ objects $\mathcal{Y}=\{\bm{Y}_1,\bm{Y}_2, \ldots,\bm{Y}_N\}$, where 
$\bm{Y}_n=  { \left[{\bm{y}^{1}_{n}},{\bm{y}^{2}_{n}} \hdots {\bm{y}^{V}_{n}} \right] }^T$, such that the following equation of triplet $(i,j,k)\in \mathcal{T}$ holds for as many triplets as possible:
\begin{align}
\text{ }\exists v\text{ s.t. }||\bm{y}^{v}_{i}-\bm{y}^{v}_{j}|| < ||\bm{y}^{v}_{i}-\bm{y}^{v}_{k}||  \label{eqMVTEOBJ}.
\end{align}
\subsubsection*{Triplet Network: }
Triplet Network~\cite{hoffer2015deep} is a deep metric learning approach for classification problems. The Triplet Network is inspired by the Siamese Network~\cite{chopra2005learning}, and it takes three samples as input and provides three embeddings corresponding to the three samples. Two of the three samples are from the same class, while the other is not. The training process for the Triplet Network minimizes the embedding distance of samples of the same class and vice-versa.
In this study, we adopt a similar approach, which takes three samples as the input for one triplet similarity comparison.

\subsubsection*{Crowd Worker Modeling: }
Some workers will recognize instances of some classes as belonging to other classes in label-annotated crowdsourcing tasks. As an alternative to combining labels using methods such as average or majority voting, a crowd layer~\cite{rodrigues2018deep} is added to automatically correct the worker bias. The crowd layer constructs confusion matrices modeling each worker, and uses an end-to-end deep learning framework to obtain their values.
Our model also uses a similar idea, constructing models for each crowd worker to represent the preference of workers. Other models that extended the crowd layer model have also been developed recently such as SpeeLFC \citep{SpeeLFC}, CoNAL \citep{conal} and LFC-x \citep{li2022beyond}.

\subsubsection*{Multiview Clustering: }
Multiview clustering methods ~\cite{zhang2021multi,yao2019multi,zhou2017randomized} typically use multiview features of a given object to perform clustering tasks. Generally, these techniques do not require raw images or texts, but instead only use the extracted features as input. In the field of image processing, features are usually obtained based on image processing techniques such as histogram of oriented gradient (HoG), local binary pattern (LBP), and scale-invariant feature transform (SIFT). Our proposed approach differs these methods in which we utilize a raw image as input, and the output is multiview embedding; then, the output can be processed with existing clustering methods.

\subsubsection*{Multi-label Learning: }
Multi-label learning~\cite{liu2021emerging,sun2010multi,ridnik2021asymmetric} is used for datasets in which a single object has multiple attributes that represent distinctive features, and each attribute corresponds to a specified label. In this study, we adopt a similar approach, which uses a shared layer structure in the neural network.

\section{Proposed Multiview Learning Framework}
In this section, we define the problem of training an end-to-end neural network that can provide multiview embeddings of given inputs and finding the view preferences of workers.~\footnote{Our implement is available at \url{https://github.com/17bit/multiview_crowdsourcing}.}
\subsection{Problem Settings}
Given the $N$-sample dataset $\mathcal{X} =\{\bm{x}_1,\bm{x}_2, \ldots, \bm{x}_N\}$, a neural network $f:\mathcal{X}\to\mathcal{Y} $ is trained to obtain $V$-view embeddings of $N$ objects $\mathcal{Y}=\{\bm{Y}_1,\bm{Y}_2, \ldots, \bm{Y}_N\}$.  

We denote by $\mathcal{T}=\{\mathcal{T}_1,\mathcal{T}_2, \ldots, \mathcal{T}_M\}$ the triplets annotated by $M$ crowd workers with multiple views, where each element in $\mathcal{T}_m$ corresponds to a \textbf{double-sided} triplet ordered tuple $(i,j,k)$, which implies that worker $m$ considers objects $i$ and $j$ among $i$, $j$, and $k$ to be more similar and can be represented by
\begin{align}
s_m(\bm{x}_i,\bm{x}_j) > s_m(\bm{x}_i,\bm{x}_k),  \\
s_m(\bm{x}_i,\bm{x}_j) > s_m(\bm{x}_j,\bm{x}_k), 
\end{align}
where
$s_m:\mathcal{X}\times \mathcal{X} \to \mathbb{R}$ is the similarity function of worker $m$. Our goal is to estimate $f$ statistically.

In some other studies~\cite{amid2015multiview,hoffer2015deep,xuan2020hard}, triplet similarity comparison was also defined as asking which of two objects $A$ and $B$ is more similar to a given object $X$. This question has only two answers; either object $A$ or $B$ must be selected. To distinguish between the two definitions, we refer to triplet comparison with two and three answers as \textbf{one-sided} and \textbf{double-sided} triplet comparison, respectively.

\subsection{Network Architecture}

\begin{figure*}[!tb]
\centerline{\includegraphics[width=6.8in]{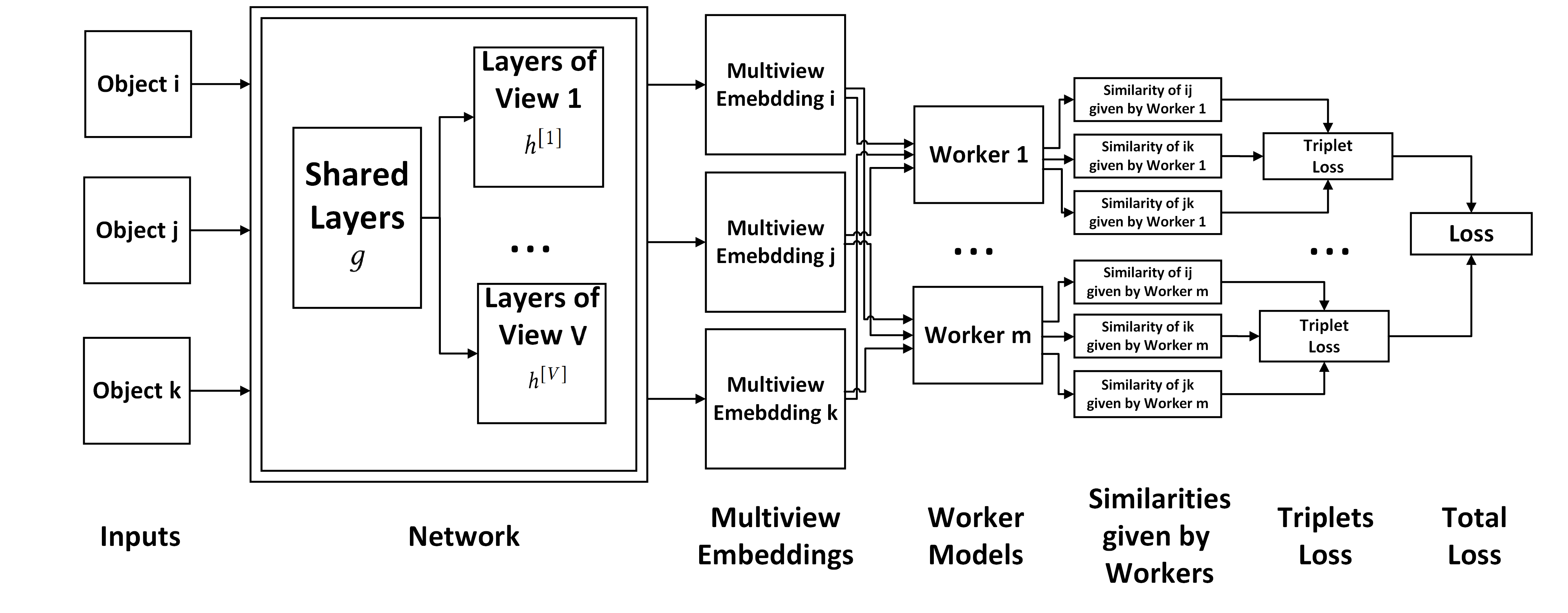}}
\caption{Architecture of multiview representation learning framework.}
\Description{Architecture of multiview representation learning framework.}
\label{arti}
\end{figure*}

Our proposed framework is shown in Figure ~\ref{arti}.
The neural network consists of two parts, including shared layers $g:\mathcal{X} \to \mathcal{Z}$ and the layers of views $h^{[1]},h^{[2]}...,h^{[V]}:\mathcal{Z} \to \mathbb{R}^D$, where $\mathcal{Z}$ is the domain of hidden embeddings of the shared layers outputs and $D$ is the number of dimensions of embeddings. The shared layers  extract common representation embeddings of objects.
Subsequent layers of views take hidden embeddings obtained from the shared layers as inputs and extract embeddings corresponding to different views to obtain multiview embeddings. The calculation process of the neural network can be written as
\begin{align}
\bm{Y}_n= f(\bm{x}_n)&=
\left[
\begin{array}{c}
    {\bm{y}^{1}_{n}}^T   \\
     {\bm{y}^{2}_{n}}^T \\
    \hdots  \\
     {\bm{y}^{V}_{n}}^T  \\  
  \end{array}
 \right] =
\left[
\begin{array}{c}
     {h^{[1]}(g(\bm{x}_n))}^T  \\
     {h^{[2]}(g(\bm{x}_n))}^T  \\
    \hdots  \\
    {h^{[V]}(g(\bm{x}_n))}^T  \\  
  \end{array}
 \right].  \label{eqYDEF}
\end{align}
Our network is based on the ResNet18 architecture~\cite{he2016deep}. The shared layers are the layers before and including the third residual block in ResNet18, and the layers of views are multiple copies of the fourth residual block and the fully connected layer.

\subsection{View Selection}
Next, we consider how to choose views in a task.
Our network outputs multiview embeddings. However, the importance of each view differs in each task. We use weights to measure their importance. If the weight of a view is large, it plays a key role in the given task.
When the triplet task is relatively simple, a view can be selected easily. For example, given three images showing "Red \textcolor{red}{O}", "Red \textcolor{red}{X}", and "Blue \textcolor{blue}{P}", a worker is more likely to note that "Red \textcolor{red}{O}" and "Red \textcolor{red}{X}" are similar because their color is the same and all the three images have different shapes. However, in more ambiguous tasks, such as the task in Figure~\ref{multi-comp}, choosing either color or dog face shape is acceptable, and depends on the preference of the workers.
Therefore, we consider that the weight can be divided into two parts, including an inherent weight and worker preference.

\subsubsection*{Inherent Weight: }
The inherent weight is only based on three objects, and we use triplet entropy to measure the inherent weight in a triplet task. If the inherent weight of a view is larger, all workers are more likely to select it in a task. Next, we provide an algorithm to calculate the triplet entropy and the inherent weight. 

Given a triplet $(i,j,k)$, the similarity function $s^v:\mathcal{X}\times \mathcal{X} \to \mathbb{R}$ between two objects in view $v$ can be defined as
\begin{align}
s^v(\bm{x}_i,\bm{x}_j) = \exp(- ||\bm{y}^v_i- \bm{y}^v_j||^2), \\
s^v(\bm{x}_i,\bm{x}_k) = \exp(- ||\bm{y}^v_i- \bm{y}^v_k||^2), \\
s^v(\bm{x}_j,\bm{x}_k) = \exp(- ||\bm{y}^v_j- \bm{y}^v_k||^2). 
\end{align}
Next, the probabilities that the similarities of one pair are the largest among three pairs can be defined as
\begin{align}
p^v_{ij\_k} = \frac{s^v(\bm{x}_i,\bm{x}_j)}{s^v(\bm{x}_i,\bm{x}_j)+s^v(\bm{x}_i,\bm{x}_k)+s^v(\bm{x}_j,\bm{x}_k)   }  \label{eqpvij},\\
p^v_{ik\_j} = \frac{s^v(\bm{x}_i,\bm{x}_k)}{s^v(\bm{x}_i,\bm{x}_j)+s^v(\bm{x}_i,\bm{x}_k)+s^v(\bm{x}_j,\bm{x}_k) },\\
p^v_{jk\_i} = \frac{s^v(\bm{x}_j,\bm{x}_k)}{s^v(\bm{x}_i,\bm{x}_j)+s^v(\bm{x}_i,\bm{x}_k)+s^v(\bm{x}_j,\bm{x}_k)  }.
\end{align}
Note that $p^v_{ij\_k} + p^v_{ik\_j} + p^v_{jk\_i} \equiv  1 $.

The triplet entropy of view $v$ can be defined as
\begin{align}
h^v_{ijk} = - ( p^v_{ij\_k}\log p^v_{ij\_k} + p^v_{ik\_j}\log p^v_{ik\_j}  +  p^v_{jk\_i}\log p^v_{jk\_i} ).
\end{align}

The maximum triplet entropy is $\log3$ when $p^v_{ij\_k} = p^v_{ik\_j} = p^v_{ik\_j} = \frac{1}{3}$ and $s^v(\bm{x}_i,\bm{x}_j) =s^v(\bm{x}_i,\bm{x}_k) = s^v(\bm{x}_j,\bm{x}_k) $, which implies that all views have equal importance and choosing between them is difficult. In contrast, the lower bound of triplet entropy is $0$, which implies that similarity between one pair is much higher than the other two, and making a choice is easy. 

Finally, the inherent weight of view $v$ of triplet task $(i,j,k)$ can be defined as the inverse of triplet entropy, that is,
\begin{align}
\widetilde{h}^v_{ijk} = \frac{\log3 - h^v_{ijk} }{\log3}, 
\end{align}
where $0\leq \widetilde{h}^v_{ijk}  < 1$.

\subsubsection*{Worker Preference: }
We denote  learnable parameters by $\mathcal{W}=\{\bm{w}_1,\bm{w}_2, \ldots, \bm{w}_M\}$, where
$\bm{w}_m=\left[
\begin{array}{cccc}
    w_m^1 & w_m^2 & \hdots & w_m^V
  \end{array}
 \right]^T,
$
the weights of $V$ views of $M$ workers. A larger value of $w_m^v$ implies that worker $m$ prefers view $v$.

\subsubsection*{Combining Weights: }
Next, we consider combining the inherent weights and worker preferences. Ideally, if a triplet task has the same inherent weights for different views, then the final weights are expected to depend on the preferences of the workers. In contrast, if a worker has no preference, then the answer is expected to depend on the inherent weights. To reflect this hypothesis, we add two weights as 
\begin{align}
q^v_{{ijk}_{(m)}}  = \widetilde{h}^v_{ijk}  + w_m^v.
\label{q}
\end{align}
We note that the initialization of $w_m^v$ follows a uniform distribution between $0$ and $1$ such that the scales of the two weights are the same.
The weights after softmax normalization can be written as
\begin{align}
\widetilde{q}^v_{{ijk}_{(m)}}  = \frac{\exp(q^v_{{ijk}_{(m)}} )}{ \sum_{v=1}^{V}{\exp(q^v_{{ijk}_{(m)}} ) }}.
\end{align}
Obviously, $\widetilde{q}^v_{{ijk}_{(m)}}=\frac{\exp( w_m^v )}{ \sum_{v=1}^{V}{\exp( w_m^v ) }}$ when $\widetilde{h}^1_{ijk}=\widetilde{h}^2_{ijk}=\cdots=\widetilde{h}^V_{ijk}$, and $\widetilde{q}^v_{{ijk}_{(m)}}=\frac{\exp( \widetilde{h}^v_{ijk})}{ \sum_{v=1}^{V}{\exp(\widetilde{h}^v_{ijk} ) }}$ when $w_m^1=w_m^2=\cdots=w_m^V$.

\subsection{Loss Function}
The similarity between two objects of triplet $(i,j,k)$ given by worker $m$ can be written as
\begin{align}
s_m(\bm{x}_i,\bm{x}_j) = \sum_{v=1}^{V} { \widetilde{q}^v_{{ijk}_{(m)}} s^v(\bm{x}_i,\bm{x}_j) }, \\
s_m(\bm{x}_i,\bm{x}_k) = \sum_{v=1}^{V} { \widetilde{q}^v_{{ijk}_{(m)}} s^v(\bm{x}_i,\bm{x}_k) }, \\
s_m(\bm{x}_j,\bm{x}_k) = \sum_{v=1}^{V} { \widetilde{q}^v_{{ijk}_{(m)}} s^v(\bm{x}_j,\bm{x}_k) }. 
\end{align}
The probability of choosing object $i$ and object $j$ as more similar by worker $m$ can be defined as
\begin{align}
p_{ij\_k_{(m)}} = \frac{s_m(\bm{x}_i,\bm{x}_j)}{s_m(\bm{x}_i,\bm{x}_j)+s_m(\bm{x}_i,\bm{x}_k)+s_m(\bm{x}_j,\bm{x}_k) }. 
\end{align}

The loss function can be derived as the following log likelihood function of all triplet tasks of all workers:
\begin{align}
\ell = -  \frac{1}{\sum_{m=1}^{M} |\mathcal{T}_m|}\sum_{m=1}^{M} \sum_{(i,j,k)\in \mathcal{T}_m} \log  p_{ij\_k_{(m)}}.
\end{align}

\section{Experiments}
First, we performed simulation experiments to verify whether our method could accurately learn the view preferences of simulated workers because knowing exactly which view a real person uses for each triplet comparison task is challenging. Subsequently, we recruited workers from the crowdsourcing platform Lancers\footnote{\url{https://www.lancers.jp}} to collect triplet data and use multiple evaluation metrics to test whether our method performed better than baseline methods.

\subsection{Datasets}
We constructed a 10-color MNIST dataset by selecting 2000 images from the MNIST~\cite{lecun-mnisthandwrittendigit-2010} dataset and using 1000 of them as the training set and the other 1000 as the test set. We chose 10 colors from the 12-color wheel\footnote{\url{https://www.usability.gov/how-to-and-tools/methods/color-basics.html}} by removing Red-orange and Yellow-orange and using a single color in each image, there were 1000 images each in the training and test sets, 100 images for each number category (from 0 to 9), and 100 images for each color category (Red, Orange, Yellow, Yellow-green, Green, Blue-green, Blue, Blue-purple, Purple and Red-purple). There were a total of 100 (number, color) pairs with 10 images in each pair. In the following experiments, we considered two images to belong to the same category if their numbers and colors were both the same.

The Stanford dog dataset~\cite{konkle2010conceptual} was used in the crowdsourcing experiments. The dataset contains 20580 images of dogs from 120 different breeds. We prepared a subset of 1000 images from 20 breeds. There were 50 images in each breed category, half of which were used for the training set and half as the test set. Therefore, there were 500 images in each of the training and test sets.

\subsection{Simulation Settings}
The data in the dog dataset could be interpreted from many uncertain hidden views, e.g., skin color, nose, eyes, and height, and finding specified criteria is challenging. Therefore, we only conducted experiments with human crowdsourcing participants for the dog dataset. For the 10-color MNIST dataset, because the data included only two possible views, we used simulated worker experiments to verify whether our proposed method could accurately find these two views. Our experiment included three different simulation settings.

In simulation setting 1, there were two workers, referred to as worker 1 and worker 2, who make decisions based on color and number, respectively. Taking the worker focusing on color as an example, he would answer if and only if the two images had the same color among three images in a triplet comparison task. Otherwise, the task is an invalid triplet query and was not included in the triplet dataset. The two workers produced 2000 triplets in each of the training and test sets.

In simulation setting 2, there were two workers, worker 1 and worker 2, who made decisions based on the distance of color and number, respectively, and the color or the number need not be exactly the same. We defined the distance of the color as the distance in the color wheel (after removing two colors), e.g., $d(\text{Red}, \text{Yellow-green})=3$. We defined the distance of the numbers as the absolute value of the difference between two numbers, e.g., $d(9, 0) = d(0, 9) = 9$. These two workers find two images with the shortest distance among three images. If two such images do not exist, the task is an invalid triplet query and was not included in the triplet dataset. The two workers produced 2000 triplets in each of the training and test sets. 

In simulation setting 3, there were four workers, of which worker 1 and worker 4 were the same as the two workers in setting 2. The remaining two workers made decisions according to the weights of $0.3$ for color, $0.7$ for number, and $0.7$ for color, $0.3$ for number, respectively. The four workers produced 1000 triplets on each of the training and the test sets, respectively. There were 4000 triplets in the training and the test sets in all the three settings. 

The difficulty of the three settings was gradually increased to verify the performance of the proposed method in different situations. Table~\ref{tabsim} gives two examples of how simulated workers choose the more similar pair among three images are given in the appendix.

\begin{table*}[!t]
\caption{Two examples for three simulation settings. The left part and the right part refers to two different triplet comparison tasks of three colored numbers. The column "$d(\text{\romannumeral 1},\text{\romannumeral 2})$" refers to the distance between the first and the second number in the triplet query and the same for the remaining columns. The row "Triplet Query" refers to triplet comparison tasks, which gives three images and asking which two of them are more similar.    }
\begin{tabular}{|ccccccccc|}
\hline
\multicolumn{1}{|c|}{~}        & \multicolumn{1}{c|}{$d(\text{\romannumeral 1},\text{\romannumeral 2})$}    & \multicolumn{1}{c|}{$d(\text{\romannumeral 1},\text{\romannumeral 3})$}    & \multicolumn{1}{c|}{$d(\text{\romannumeral 2},\text{\romannumeral 3})$}    & \multicolumn{1}{c|}{More Similar Pair} & 
\multicolumn{1}{c|}{$d(\text{\romannumeral 1},\text{\romannumeral 2})$}    & \multicolumn{1}{c|}{$d(\text{\romannumeral 1},\text{\romannumeral 3})$}    & \multicolumn{1}{c|}{$d(\text{\romannumeral 2},\text{\romannumeral 3})$}     & More Similar Pair \\ \hline
\multicolumn{1}{|c|}{Triplet Query} & \multicolumn{4}{c|}{red \textcolor{red}{1} red \textcolor{red}{2} green \textcolor{green}{2}}                                                                                                             & \multicolumn{4}{c|}{red \textcolor{red}{1} orange \textcolor{orange}{2} green \textcolor{green}{3}}                                                                                        \\ \hline
\multicolumn{9}{|c|}{Simulation Setting 1}                                                                                                                                                                                                                                                       \\ \hline
\multicolumn{1}{|c|}{Worker 1}      & \multicolumn{1}{c|}{Same}     & \multicolumn{1}{c|}{Not Same} & \multicolumn{1}{c|}{Not Same} & \multicolumn{1}{c|}{red \textcolor{red}{1} and red \textcolor{red}{2}}             & \multicolumn{1}{c|}{Not Same} & \multicolumn{1}{c|}{Not Same} & \multicolumn{1}{c|}{Not Same} & Invalid Query     \\ \hline
\multicolumn{1}{|c|}{Worker 2}      & \multicolumn{1}{c|}{Not Same} & \multicolumn{1}{c|}{Not Same} & \multicolumn{1}{c|}{Same}     & \multicolumn{1}{c|}{red \textcolor{red}{2} and green \textcolor{green}{2}}             & \multicolumn{1}{c|}{Not Same} & \multicolumn{1}{c|}{Not Same} & \multicolumn{1}{c|}{Not Same} & Invalid Query     \\ \hline
\multicolumn{9}{|c|}{Simulation Setting 2}                                                                                                                                                                                                                                                       \\ \hline
\multicolumn{1}{|c|}{Worker 1}      & \multicolumn{1}{c|}{0}        & \multicolumn{1}{c|}{4}        & \multicolumn{1}{c|}{4}        & \multicolumn{1}{c|}{red \textcolor{red}{1} and red \textcolor{red}{2}}             & \multicolumn{1}{c|}{1}        & \multicolumn{1}{c|}{4}        & \multicolumn{1}{c|}{3}        & red \textcolor{red}{1} and orange \textcolor{orange}{2}            \\ \hline
\multicolumn{1}{|c|}{Worker 2}      & \multicolumn{1}{c|}{1}        & \multicolumn{1}{c|}{1}        & \multicolumn{1}{c|}{0}        & \multicolumn{1}{c|}{red \textcolor{red}{2} and green \textcolor{green}{2}}             & \multicolumn{1}{c|}{1}        & \multicolumn{1}{c|}{2}        & \multicolumn{1}{c|}{1}        & Invalid Query     \\ \hline
\multicolumn{9}{|c|}{Simulation Setting 3}                                                                                                                                                                                                                                                       \\ \hline
\multicolumn{1}{|c|}{Worker 1}      & \multicolumn{1}{c|}{0}        & \multicolumn{1}{c|}{4}        & \multicolumn{1}{c|}{4}        & \multicolumn{1}{c|}{red \textcolor{red}{1} and red \textcolor{red}{2}}             & \multicolumn{1}{c|}{1}        & \multicolumn{1}{c|}{4}        & \multicolumn{1}{c|}{3}        & red \textcolor{red}{1} and orange \textcolor{orange}{2}             \\ \hline
\multicolumn{1}{|c|}{Worker 2}      & \multicolumn{1}{c|}{0.3}      & \multicolumn{1}{c|}{3.1}      & \multicolumn{1}{c|}{2.8}      & \multicolumn{1}{c|}{red \textcolor{red}{1} and red \textcolor{red}{2}}             & \multicolumn{1}{c|}{1}        & \multicolumn{1}{c|}{3.4}      & \multicolumn{1}{c|}{2.4}      & red \textcolor{red}{1} and orange \textcolor{orange}{2}            \\ \hline
\multicolumn{1}{|c|}{Worker 3}      & \multicolumn{1}{c|}{0.7}      & \multicolumn{1}{c|}{1.9}      & \multicolumn{1}{c|}{1.2}      & \multicolumn{1}{c|}{red \textcolor{red}{1} and red \textcolor{red}{2}}             & \multicolumn{1}{c|}{1}        & \multicolumn{1}{c|}{2.6}      & \multicolumn{1}{c|}{1.6}      & red \textcolor{red}{1} and orange \textcolor{orange}{2}            \\ \hline
\multicolumn{1}{|c|}{Worker 4}      & \multicolumn{1}{c|}{1}        & \multicolumn{1}{c|}{1}        & \multicolumn{1}{c|}{0}        & \multicolumn{1}{c|}{red \textcolor{red}{2} and green \textcolor{green}{2}}             & \multicolumn{1}{c|}{1}        & \multicolumn{1}{c|}{2}        & \multicolumn{1}{c|}{1}        & Invalid Query     \\ \hline
\end{tabular}
\label{tabsim}
\end{table*}

\subsection{Real Crowdsourcing Setting}
\begin{figure}[t]
 \centerline{\includegraphics[width=2.5in]{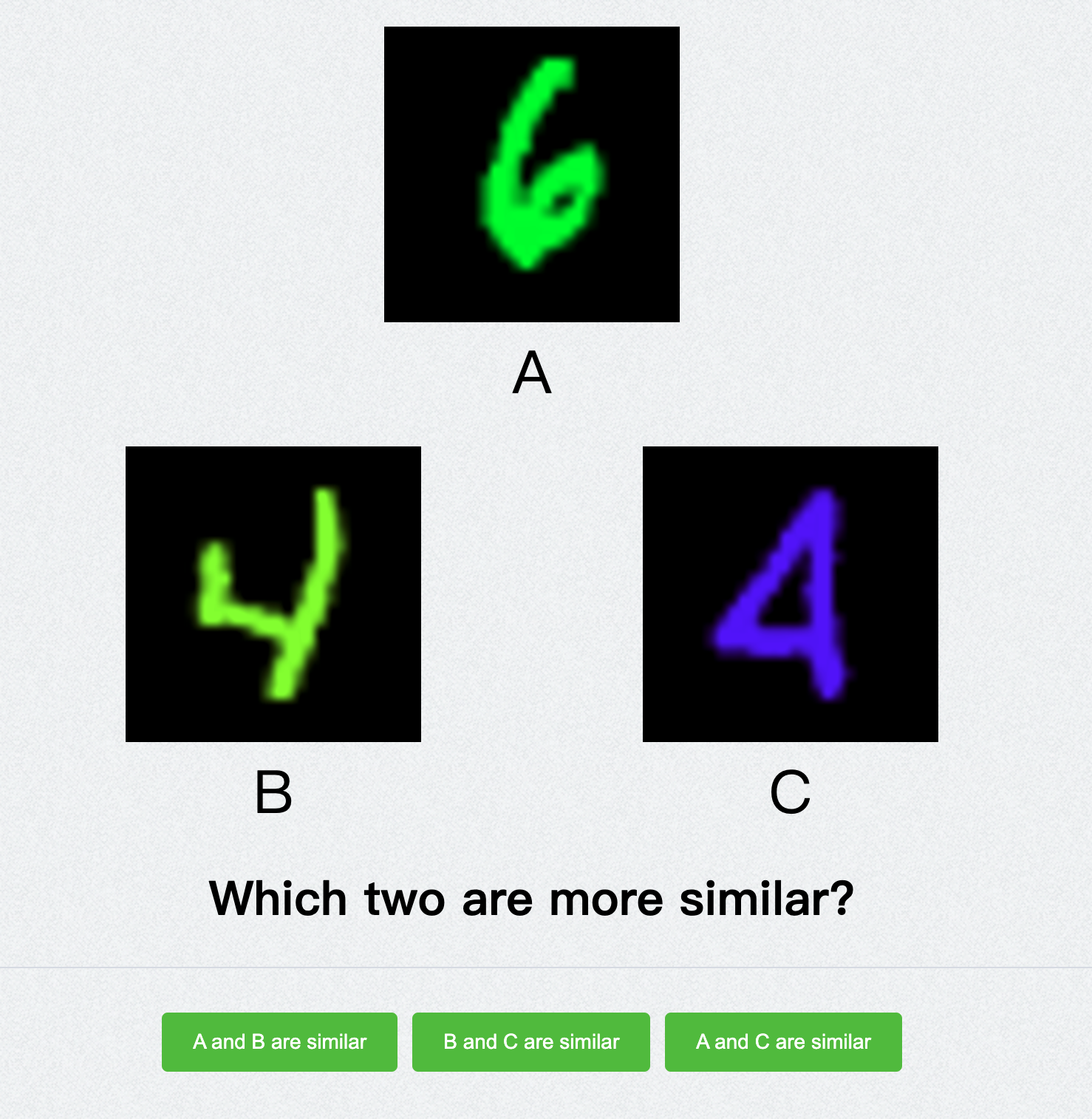}}
 \caption{The crowdsourcing task of triplet similarity comparison. Three images are put in a triangle shape to avoid the distance between them affecting decisions of crowd workers.}
 \Description{The crowdsourcing task of triplet similarity comparison.} \label{crowdplat}
 \end{figure}

80 workers were recruited for both the 10-color MNIST dataset and the dog dataset, respectively. Each worker was required to complete 100 random triplet comparison tasks shown in Fig.~\ref{crowdplat} for a reward of JPY220 on the Lancers platform, of which images of 50 tasks were used from the training set and the other 50 tasks were used from the test set. We did not filter the data to avoid bias, except for workers who completed the tasks within an extremely short period and that implied obvious suspicions of submissions of inferior quality for the tasks. Workers completed the task in roughly 10 minutes on average. The hourly pay was about JPY1300, which exceeds the minimum hourly wage, about JPY1000, in Japan. We consider that if the choice of a given worker is inconsistent with that of the majority, this was because their view was in the minority, not because their choice is wrong.

\subsection{Evaluation Metrics}
Accuracy refers to the proportion of cases in which the network can successfully give the correct pair highest similarity among three objects for the triplet data. 
We use triplet data to train the network, but our goal was to train a neural network $f$ to obtain multiview embeddings, which means that we need to evaluate the performance of embeddings $\mathcal{Y}$ without using triplet data $\mathcal{T}$. Therefore, the accuracy of the test set was not applicable. Several different evaluation metrics can be considered. 

\subsubsection*{Clustering Evaluation: }
We clustered the embeddings of the test set without labels and evaluated the results. In our experiments, we used $k$-means and agglomerative clustering methods as well as purity and normalized mutual information (NMI) evaluation metrics. 

\subsubsection*{Linear Evaluation: }
We evaluated the performance of representation learning models in fine-tuning tasks~\cite{chen2020simple, zhang2016colorful,oord2018representation}. A linear layer was trained by the embeddings and ground truth labels of the training set. Then, the linear layer was applied to the embeddings of the test set to predict the labels. The accuracy of classification was the evaluation score.

\subsubsection*{$K$-anchors Evaluation: }
We evaluated the performance of predicting labels when only labels of a small number of samples were known.
$K$ samples were randomly selected from each category as anchors in the test set. We then used anchors to predict labels in the test set. The prediction of each sample was the category of a selected anchor sample with the minimum Euclidean distance. The accuracy of classification was the evaluation score.

\subsection{Experimental Results}

Our proposed methods with $2$ views performed better than the baselines on both simulation experiments and real crowdsourcing experiments in all the evaluation metrics, as shown in Tables~\ref{tab21} and \ref{tab22} respectively.
Experiments with a single view were used as baseline, and the neural network architecture was exactly the same as the common ResNet18. The results show that compared to the single view,  setting the number of views to $2$ lead to a significant improvement.

Next, we discuss cases in which we used more views. Table~\ref{tab23} indicates that increasing the number of views improved performance slightly, but not as much as introducing multiview, which increased the number of views from 1 to 2.

\begin{table*}[!t]

\caption{Results on simulation experiments on the 10-color MINST test set. A larger value indicates better performance. }

\centering

\begin{tabular}{|cccccccc|}
\hline
\multicolumn{1}{|c|}{Number of views}            & \multicolumn{1}{c|}{Accuracy}        & \multicolumn{4}{c|}{Clustering Eval.}                                                                                                                     & \multicolumn{1}{c|}{Linear Eval.}    & $2$-anchors Eval. \\ \hline
\multicolumn{2}{|c|}{\multirow{2}{*}{}}                                       & \multicolumn{2}{c|}{K-means}                                                & \multicolumn{2}{c|}{Agglomerative}                                          & \multicolumn{2}{c|}{\multirow{2}{*}{}}                   \\ \cline{3-6}
\multicolumn{2}{|c|}{}                                                        & \multicolumn{1}{c|}{Purity}          & \multicolumn{1}{c|}{NMI}             & \multicolumn{1}{c|}{Purity}          & \multicolumn{1}{c|}{NMI}             & \multicolumn{2}{c|}{}                \\ \hline
\multicolumn{8}{|c|}{Simulation Setting 1 on the 10-color MNIST dataset}                                                                                                                                                                                                                                                           \\ \hline
\multicolumn{1}{|c|}{1 (Baseline)}                & \multicolumn{1}{c|}{0.7370}          & \multicolumn{1}{c|}{0.7420}          & \multicolumn{1}{c|}{0.8941}          & \multicolumn{1}{c|}{0.7595}          & \multicolumn{1}{c|}{0.9041}          & \multicolumn{1}{c|}{0.8631}          & 0.6989            \\ \hline
\multicolumn{1}{|c|}{\textbf{2 (Ours)}} & \multicolumn{1}{c|}{\textbf{0.9750}} & \multicolumn{1}{c|}{\textbf{0.8990}} & \multicolumn{1}{c|}{\textbf{0.9430}} & \multicolumn{1}{c|}{\textbf{0.9092}} & \multicolumn{1}{c|}{\textbf{0.9525}} & \multicolumn{1}{c|}{\textbf{0.9877}} & \textbf{0.9033}   \\ \hline
\multicolumn{8}{|c|}{Simulation Setting 2 on the 10-color MNIST dataset}                                                                                                                                                                                                                                                           \\ \hline
\multicolumn{1}{|c|}{1 (Baseline)}                & \multicolumn{1}{c|}{0.5745}          & \multicolumn{1}{c|}{0.3523}          & \multicolumn{1}{c|}{0.6560}          & \multicolumn{1}{c|}{0.3479}          & \multicolumn{1}{c|}{0.6584}          & \multicolumn{1}{c|}{0.6491}          & 0.4855            \\ \hline
\multicolumn{1}{|c|}{\textbf{2 (Ours)}} & \multicolumn{1}{c|}{\textbf{0.8825}} & \multicolumn{1}{c|}{\textbf{0.6001}} & \multicolumn{1}{c|}{\textbf{0.8244}} & \multicolumn{1}{c|}{\textbf{0.6282}} & \multicolumn{1}{c|}{\textbf{0.8290}} & \multicolumn{1}{c|}{\textbf{0.9522}} & \textbf{0.7413}   \\ \hline
\multicolumn{8}{|c|}{Simulation Setting 3 on the 10-color MNIST dataset}                                                                                                                                                                                                                                                       \\ \hline
\multicolumn{1}{|c|}{1 (Baseline)}                & \multicolumn{1}{c|}{0.6002}          & \multicolumn{1}{c|}{0.3772}          & \multicolumn{1}{c|}{0.6659}          & \multicolumn{1}{c|}{0.3570}          & \multicolumn{1}{c|}{0.6598}          & \multicolumn{1}{c|}{0.6691}          & 0.4549            \\ \hline
\multicolumn{1}{|c|}{\textbf{2 (Ours)}} & \multicolumn{1}{c|}{\textbf{0.8051}} & \multicolumn{1}{c|}{\textbf{0.5792}} & \multicolumn{1}{c|}{\textbf{0.7861}} & \multicolumn{1}{c|}{\textbf{0.5958}} & \multicolumn{1}{c|}{\textbf{0.8091}} & \multicolumn{1}{c|}{\textbf{0.9151}} & \textbf{0.7381}   \\ \hline
\end{tabular}

\label{tab21}
\end{table*}

\begin{table*}[!t]

\caption{Results on real crowdsourcing experiments on the 10-color MINST and Dog test sets. }

\centering
\begin{tabular}{|cccccccc|}
\hline
\multicolumn{1}{|c|}{Number of views}            & \multicolumn{1}{c|}{Accuracy}        & \multicolumn{4}{c|}{Clustering Eval.}                                                                                                                     & \multicolumn{1}{c|}{Linear Eval.}    & $2$-anchors Eval. \\ \hline
\multicolumn{2}{|c|}{\multirow{2}{*}{}}                                       & \multicolumn{2}{c|}{K-means}                                                & \multicolumn{2}{c|}{Agglomerative}                                          & \multicolumn{2}{c|}{\multirow{2}{*}{}}                   \\ \cline{3-6}
\multicolumn{2}{|c|}{}                                                        & \multicolumn{1}{c|}{Purity}          & \multicolumn{1}{c|}{NMI}             & \multicolumn{1}{c|}{Purity}          & \multicolumn{1}{c|}{NMI}             & \multicolumn{2}{c|}{}                \\ \hline
\multicolumn{8}{|c|}{10-color MNIST}                                                                                                                                                                                                                                                                 \\ \hline
\multicolumn{1}{|c|}{1 (Baseline)}                & \multicolumn{1}{c|}{0.4583}          & \multicolumn{1}{c|}{0.2858}          & \multicolumn{1}{c|}{0.5041}          & \multicolumn{1}{c|}{0.3062}          & \multicolumn{1}{c|}{0.5291}          & \multicolumn{1}{c|}{0.6701}          & 0.3375            \\ \hline
\multicolumn{1}{|c|}{\textbf{2 (Ours)}} & \multicolumn{1}{c|}{\textbf{0.6941}} & \multicolumn{1}{c|}{\textbf{0.5461}} & \multicolumn{1}{c|}{\textbf{0.7539}} & \multicolumn{1}{c|}{\textbf{0.5587}} & \multicolumn{1}{c|}{\textbf{0.7784}} & \multicolumn{1}{c|}{\textbf{0.8243}} & \textbf{0.5899}   \\ \hline
\multicolumn{8}{|c|}{Dog}                                                                                                                                                                                                                                                                            \\ \hline
\multicolumn{1}{|c|}{1 (Baseline)}                & \multicolumn{1}{c|}{0.3451}          & \multicolumn{1}{c|}{0.1831}          & \multicolumn{1}{c|}{0.3132}          & \multicolumn{1}{c|}{0.1697}          & \multicolumn{1}{c|}{0.3302}          & \multicolumn{1}{c|}{0.4892}          & 0.1344            \\ \hline
\multicolumn{1}{|c|}{\textbf{2 (Ours)}} & \multicolumn{1}{c|}{\textbf{0.4882}} & \multicolumn{1}{c|}{\textbf{0.3122}} & \multicolumn{1}{c|}{\textbf{0.6402}} & \multicolumn{1}{c|}{\textbf{0.3351}} & \multicolumn{1}{c|}{\textbf{0.6544}} & \multicolumn{1}{c|}{\textbf{0.5444}} & \textbf{0.2941}   \\ \hline
\end{tabular}

\label{tab22}
\end{table*}

\begin{table*}[!t]

\caption{Results of different number of views on the Dog test set.  }

\centering

\begin{tabular}{|cc|cccc|cc|}
\hline
\multicolumn{1}{|c|}{Number of views}  & Accuracy & \multicolumn{4}{c|}{Clustering Eval.}                                                            & \multicolumn{1}{c|}{Linear Eval.} & $2$-anchors Eval. \\ \hline
\multicolumn{2}{|c|}{\multirow{2}{*}{}} & \multicolumn{2}{c|}{K-means}                              & \multicolumn{2}{c|}{Agglomerative}   & \multicolumn{2}{c|}{\multirow{2}{*}{}}                \\ \cline{3-6}
\multicolumn{2}{|c|}{}                  & \multicolumn{1}{c|}{Purity} & \multicolumn{1}{c|}{NMI}    & \multicolumn{1}{c|}{Purity} & NMI    & \multicolumn{2}{c|}{}             \\ \hline
\multicolumn{1}{|c|}{2}      & 0.4882   & \multicolumn{1}{c|}{0.3122} & \multicolumn{1}{c|}{0.6402} & \multicolumn{1}{c|}{0.3351} & 0.6544 & \multicolumn{1}{c|}{0.5444}       & 0.2941            \\ \hline
\multicolumn{1}{|c|}{3}      & 0.5041   & \multicolumn{1}{c|}{0.3503} & \multicolumn{1}{c|}{0.6395} & \multicolumn{1}{c|}{0.3683} & 0.6504 & \multicolumn{1}{c|}{0.5693}       & 0.3310            \\ \hline
\multicolumn{1}{|c|}{4}      & 0.5191   & \multicolumn{1}{c|}{0.3641} & \multicolumn{1}{c|}{0.6485} & \multicolumn{1}{c|}{0.3684} & 0.6528 & \multicolumn{1}{c|}{0.5592}       & 0.3593            \\ \hline
\end{tabular}

\label{tab23}
\end{table*}

\begin{table}[!t]

\caption{Results of how much workers prefer view 1 on simulation experiments on 10-color MNIST. The value of worker $m$ is defined as $\frac{\exp(w_m^1)}{\exp(w_m^1)+\exp(w_m^2)}$. A larger value indicates that the worker prefers view 1. }
\centering
 \resizebox{1.0\linewidth}{!}{
\begin{tabular}{c|cccc}
\toprule
{\small Simulation Setting} & 
Worker 1 & 
Worker 2  & 
Worker 3  & 
Worker 4  \\ \midrule
1   & 0.9799   & 0.0695  & N/A    & N/A     \\
2     & 0.0501  & 0.9493       & N/A     & N/A     \\
3     & 0.1103  & 0.4281       & 0.7833       & 0.9514 \\
\bottomrule
\end{tabular}
}
\label{tab3}
\end{table}

\begin{figure*}[!t]
\begin{tabular}{ccccc}
 ~ &
 \textbf{Number of Views = $1$} &
 \textbf{Number of Views = $2$ Global} &
 \textbf{Number of Views = $2$ View 1} &
 \textbf{Number of Views = $2$ View 2} \\

 \rotatebox{90}{\textbf{Simulation Setting 1}} &
 \includegraphics[width=1.45in]{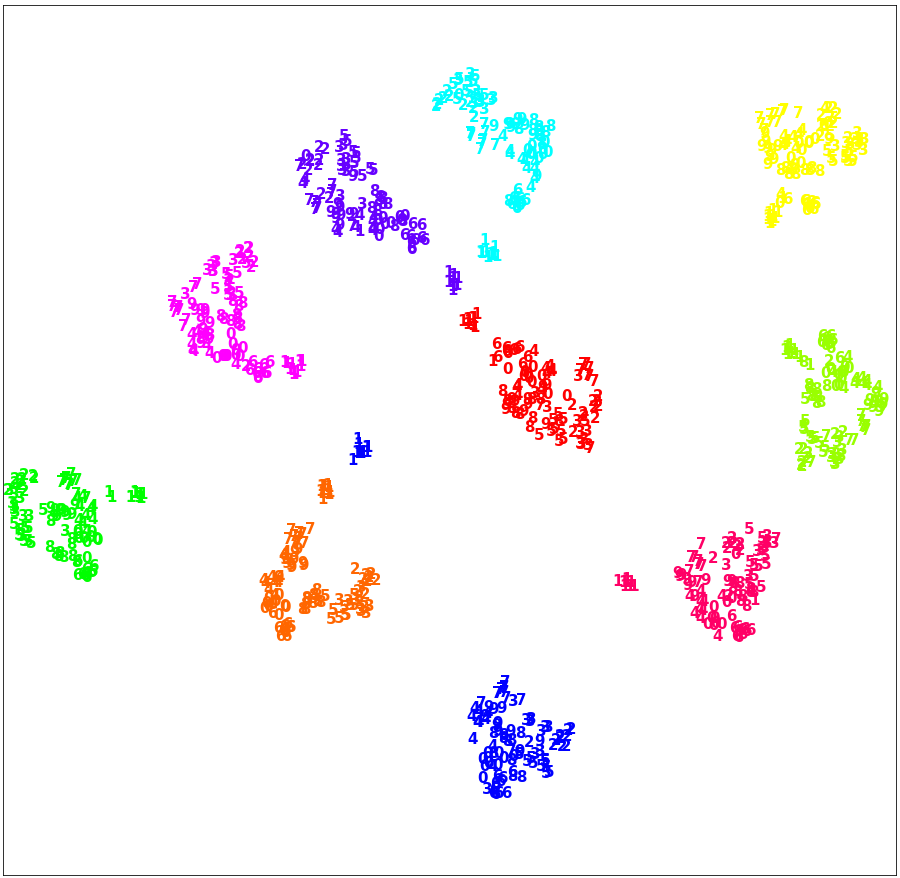}  &
 \includegraphics[width=1.45in]{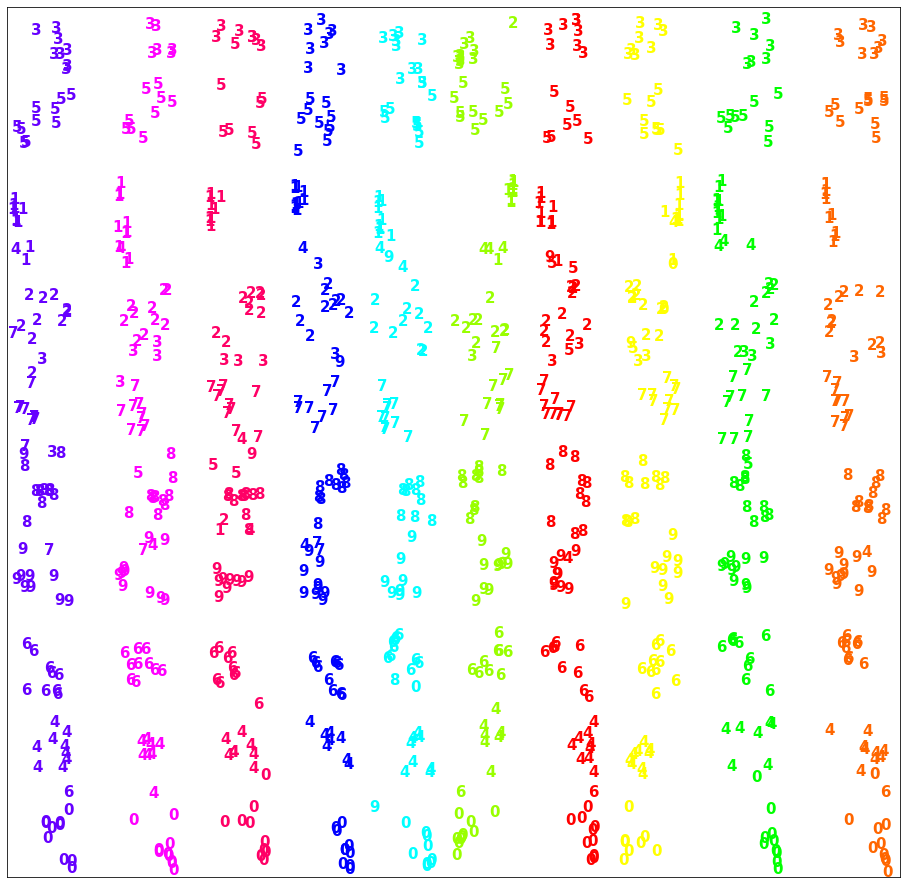}  &
 \includegraphics[width=1.45in]{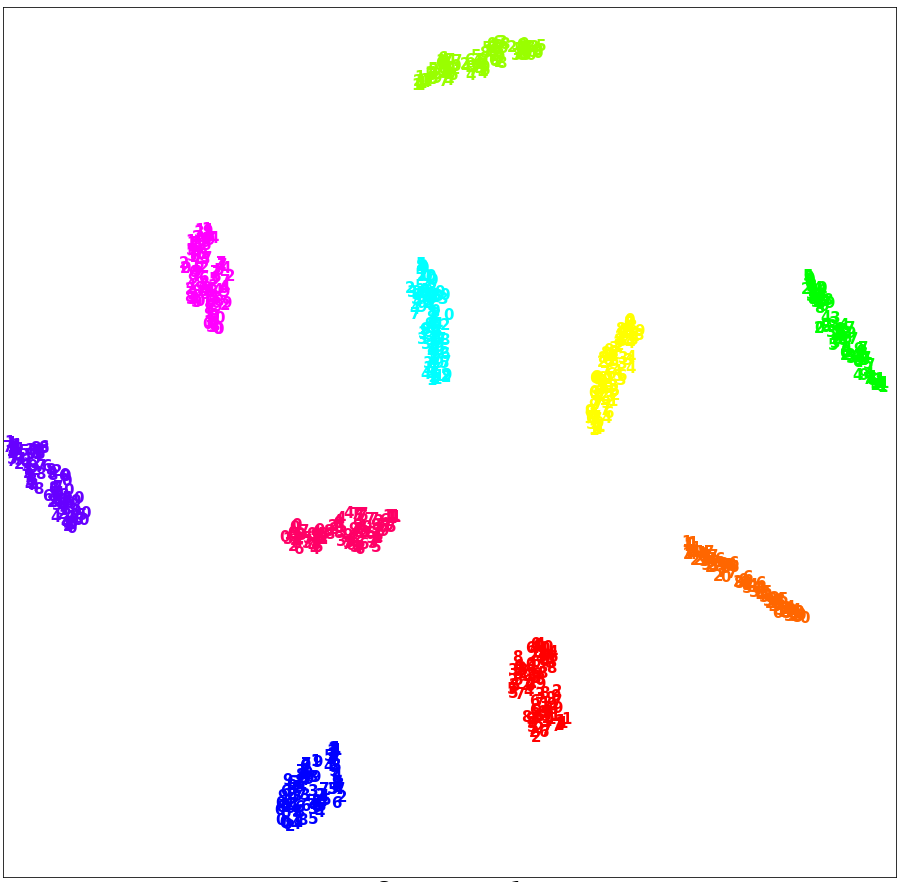}  &
 \includegraphics[width=1.45in]{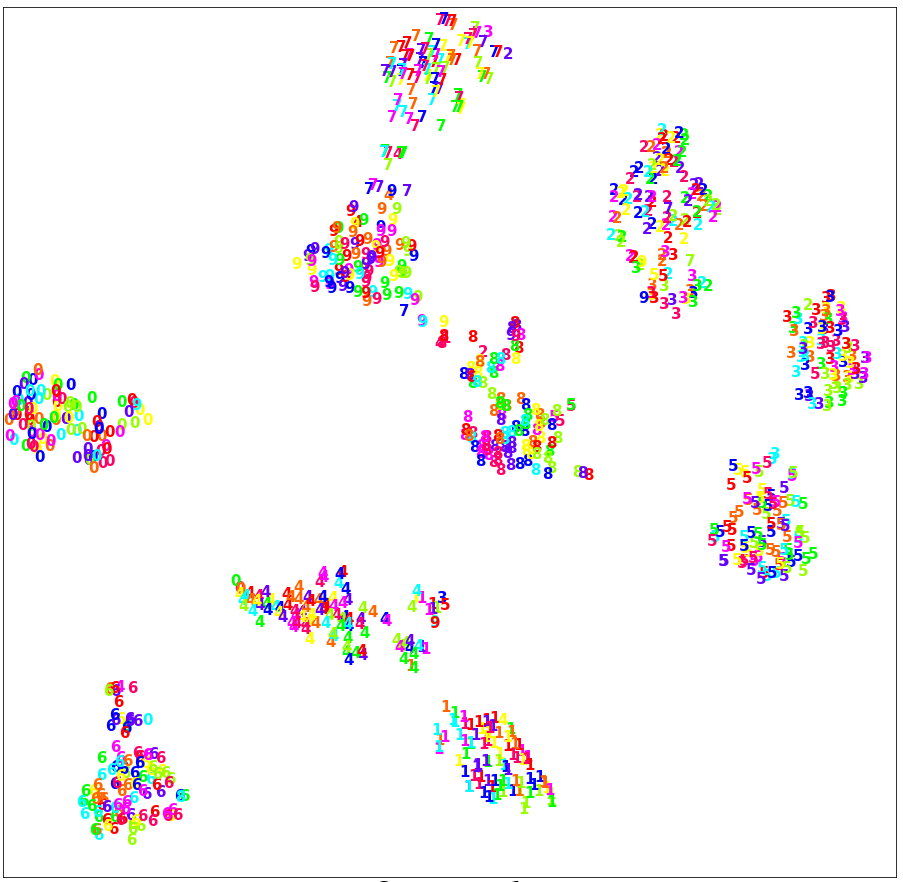}  
 \\

 \rotatebox{90}{~\textbf{Simulation Setting 2}} &
 \includegraphics[width=1.45in]{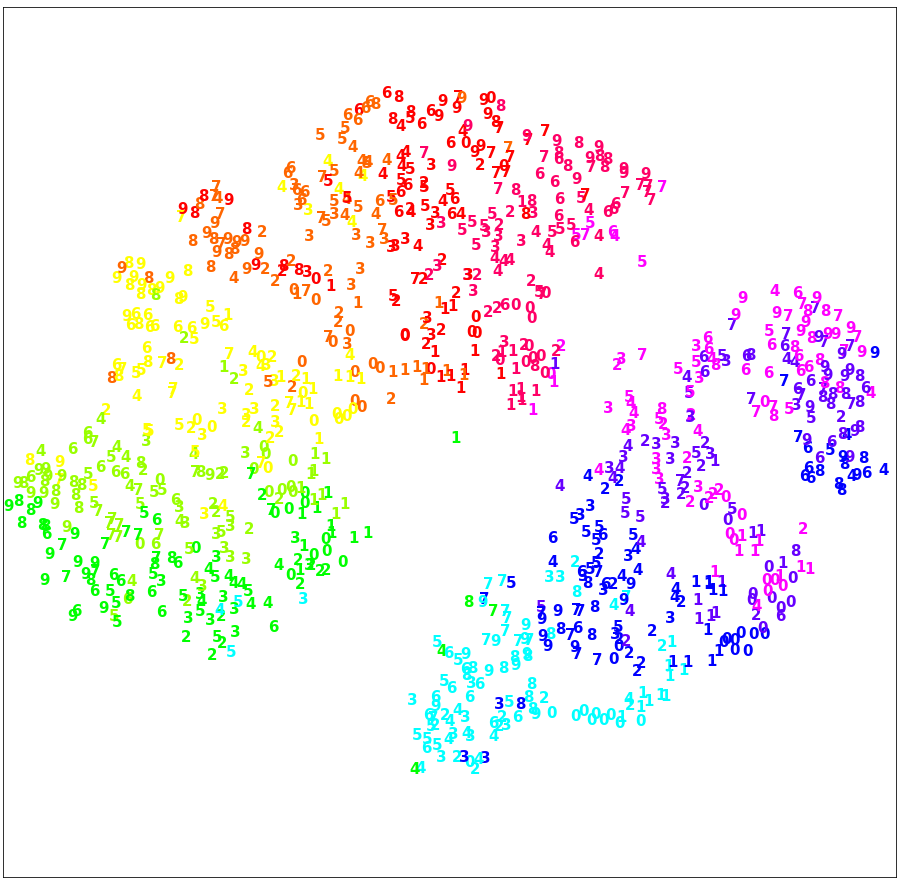}  &
 \includegraphics[width=1.45in]{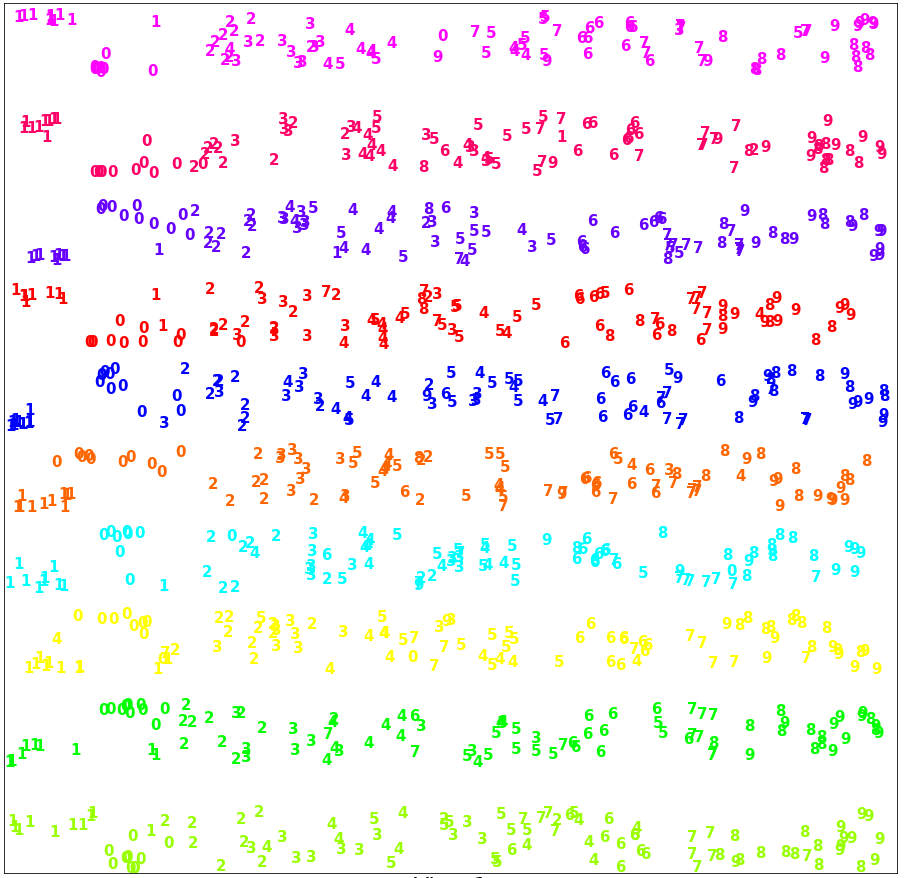}  &
 \includegraphics[width=1.45in]{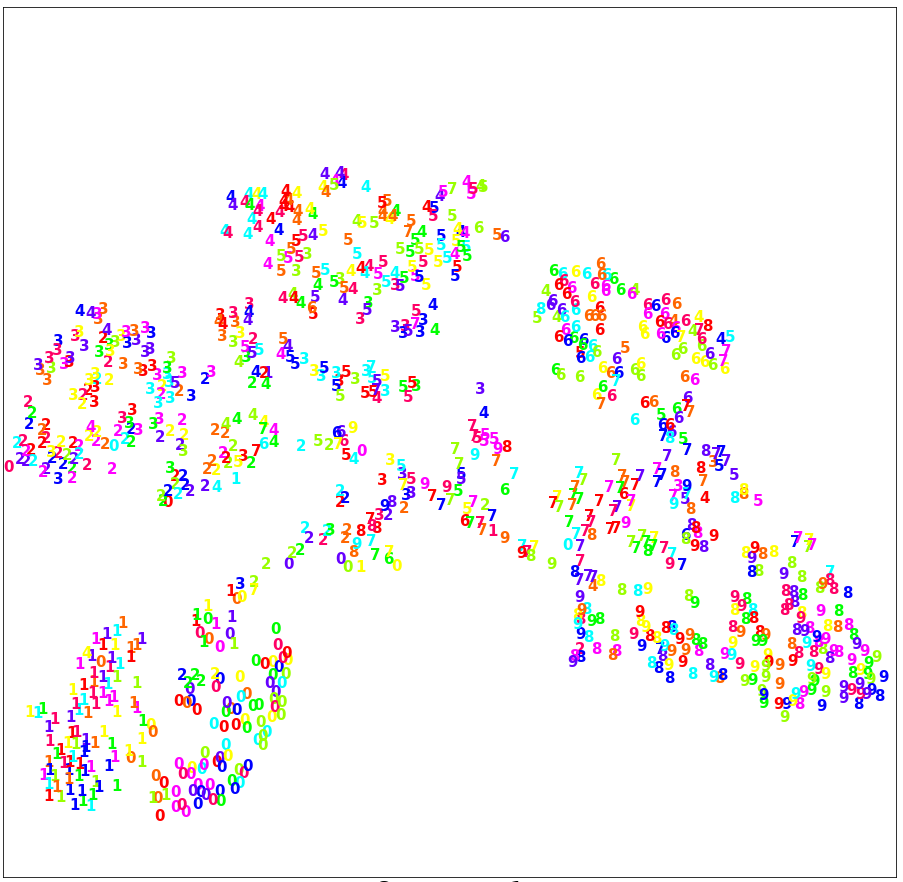}  &
 \includegraphics[width=1.45in]{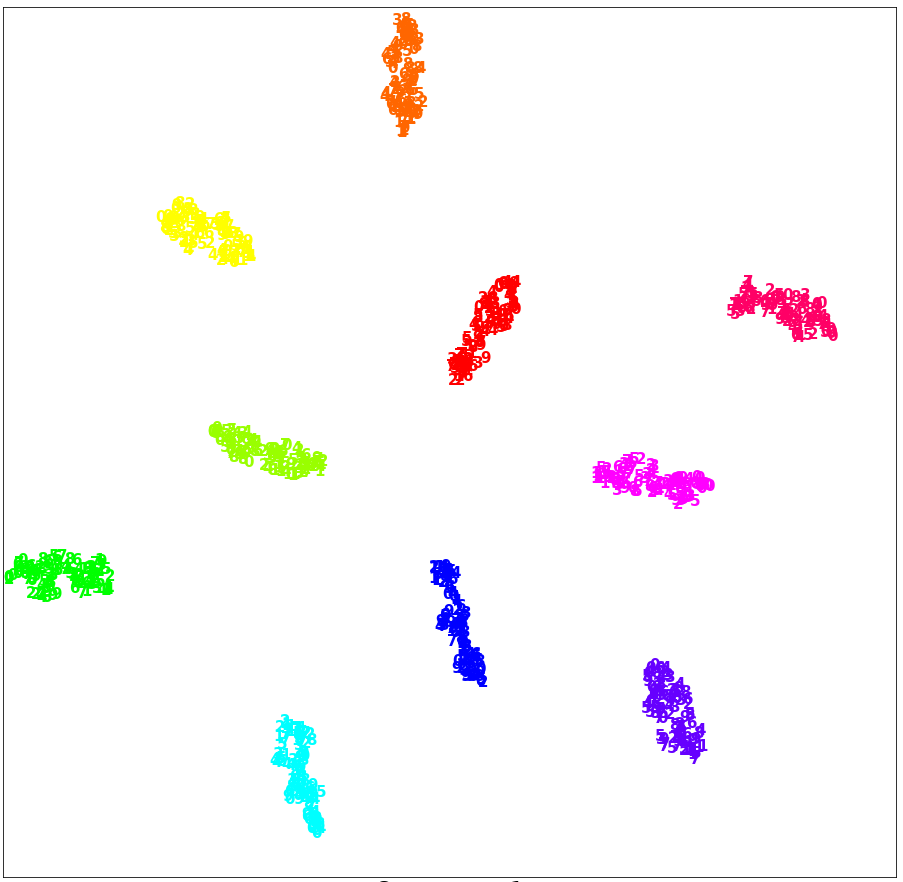}  
 \\
 
  \rotatebox{90}{~\textbf{Simulation Setting 3}} &
 \includegraphics[width=1.45in]{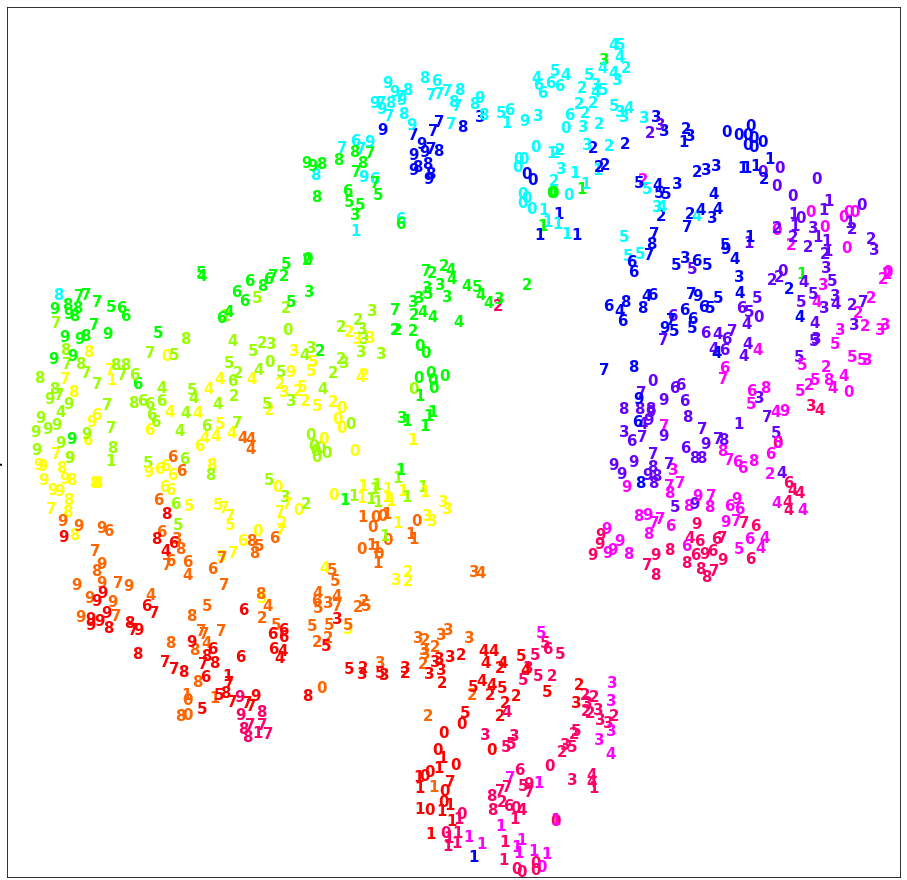}  &
 \includegraphics[width=1.45in]{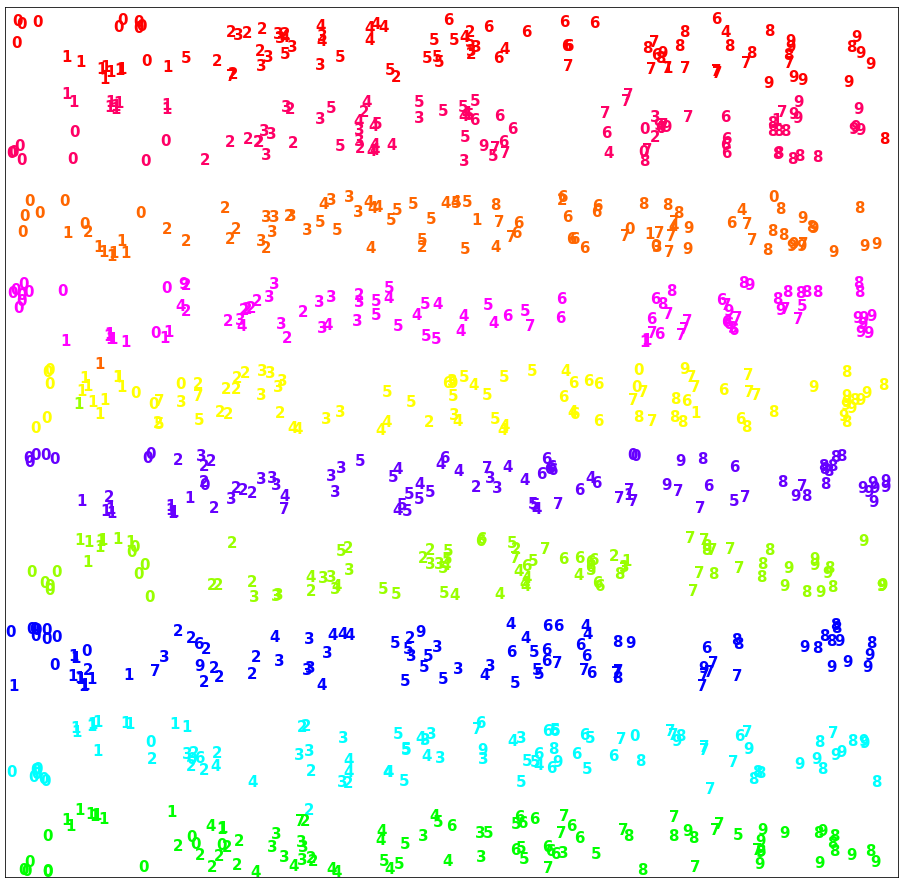}  &
 \includegraphics[width=1.45in]{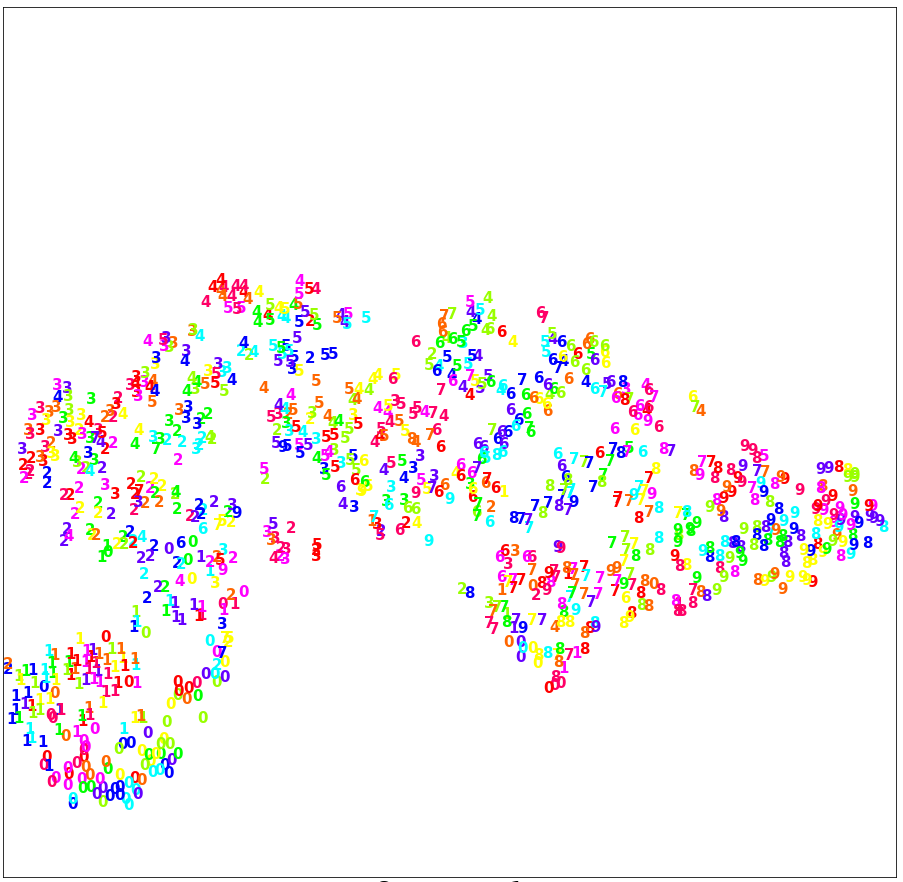}  &
 \includegraphics[width=1.45in]{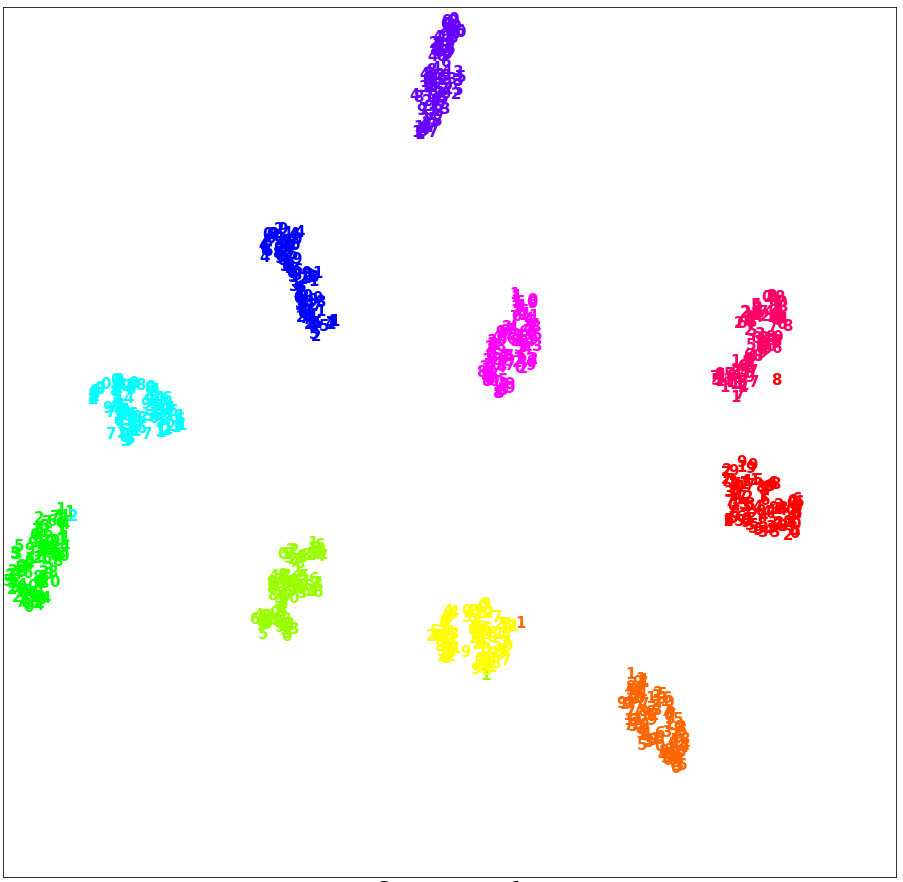}  
 \\
 
  \rotatebox{90}{\textbf{Real Crowdsourcing}} &
 \includegraphics[width=1.45in]{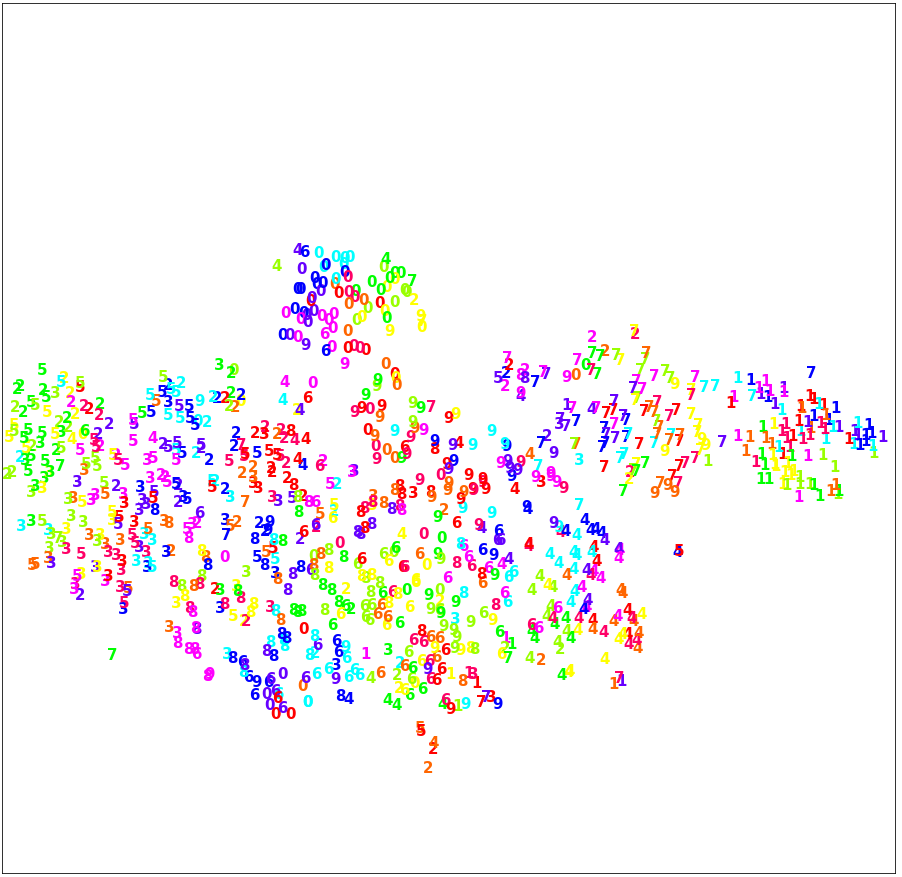}  &
 \includegraphics[width=1.45in]{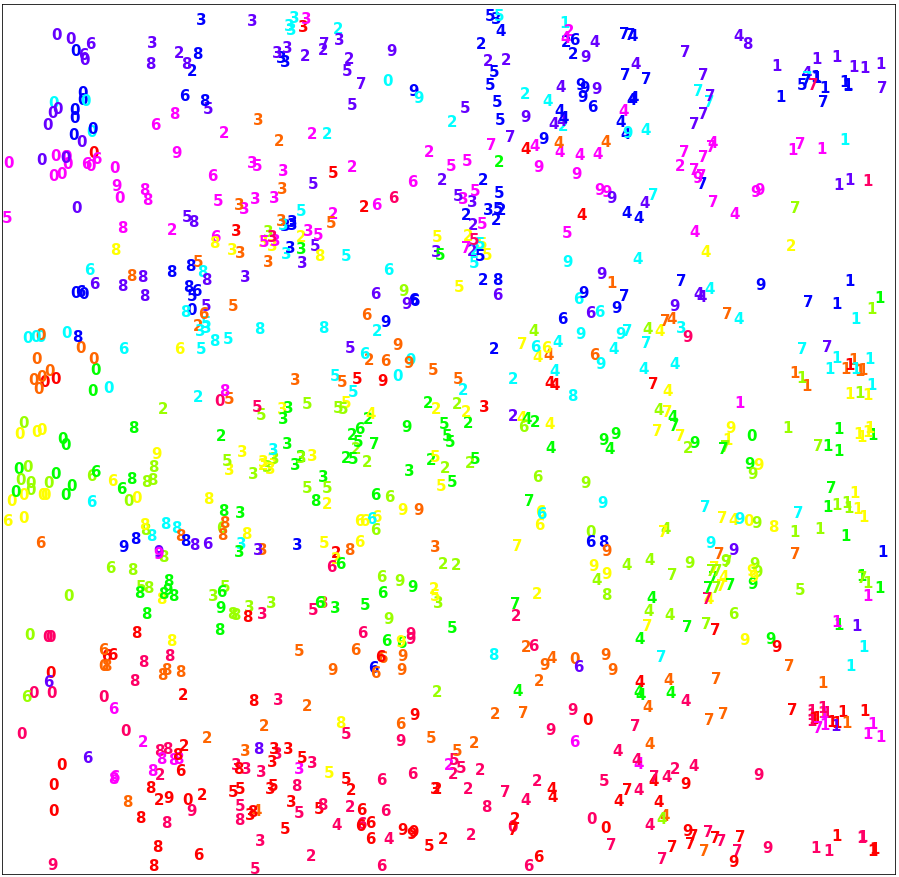}  &
 \includegraphics[width=1.45in]{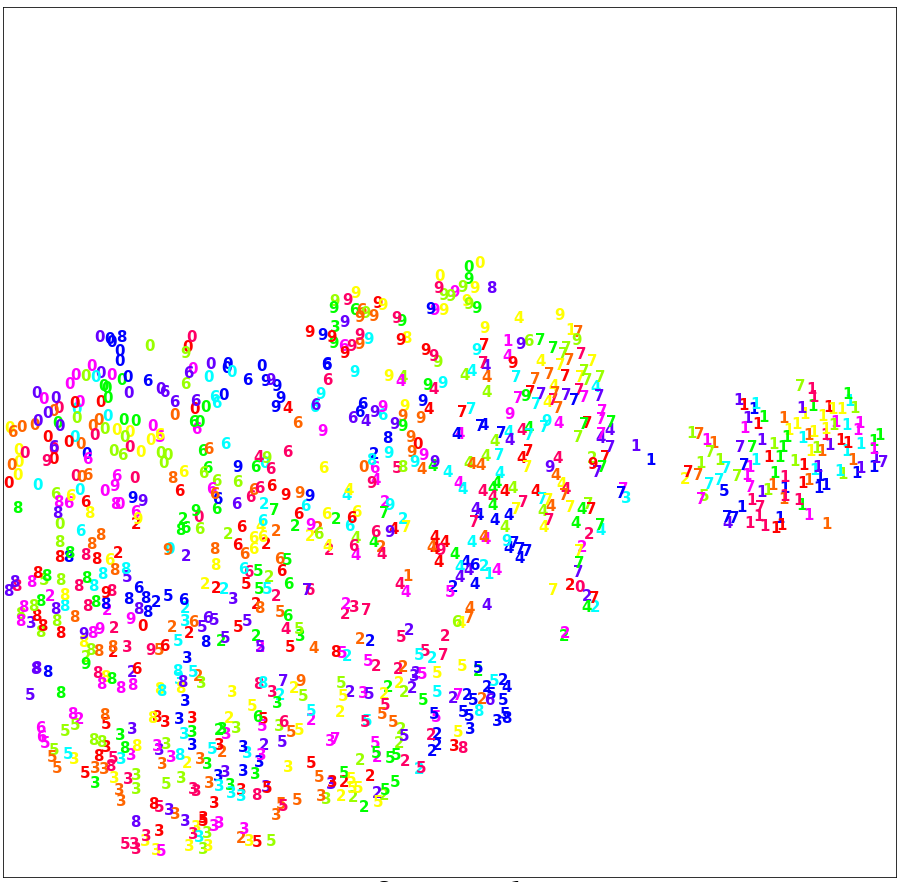}  &
 \includegraphics[width=1.45in]{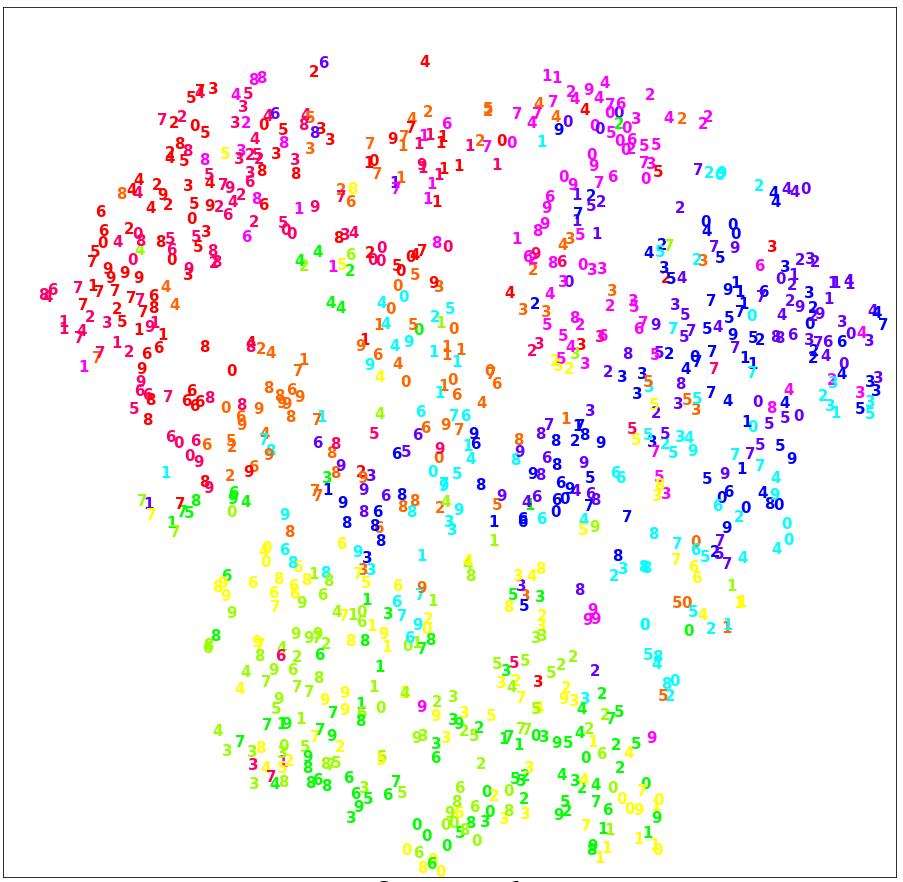}  
 \\

\end{tabular}

\caption{t-SNE visualization of embeddings from the 10-color MNIST dataset. Each number in the figure corresponds to a sample in the test set (Zoom in to see colored numbers clearly). The second column shows the global visualization of the experiment with two views, where the third and fourth columns were Views 1 and 2 respectively. 
For the second column, the dimensions of embeddings of each view were reduced from $D$ to $1$. The horizontal and vertical axes indicate Views 1 and 2, respectively. For the first, the third and the fourth column, the dimensions of embeddings were reduced from $D$ to $2$. The horizontal and vertical axes indicate component 1 and component 2, respectively.  } 
\Description{t-SNE visualization of embeddings from the 10-color MNIST dataset.}
\label{tsne}
\end{figure*}

Figure~\ref{tsne} shows the results of visualization using t-SNE~\cite{van2008visualizing} for the 10-color MNIST dataset. 
When the number of views was $1$, it shows that samples were clustered by color in all the three simulation setting, and by number in real crowdsourcing experiments. None were clustered by color and shape at the same time. However, views 1 and 2 corresponded to color and number respectively in simulation setting 1 and to number and color respectively in simulation setting 2, simulation setting 3, and the real crowdsourcing experiments.
The embeddings also learned the distance relationships given by workers. For example, it may be observed that numbers were roughly ordered from 0 to 9 in view 1 (see the third row and the third column in Figure~\ref{tsne})  and the color distribution roughly matched the 12-color wheel in view 2 (see the third row and the fourth column in Figure~\ref{tsne}) in simulation setting 3. 
It may also be observed that 0s (samples of 0 in the figure) and 8s were close, whereas 1s and 7s are close in the real crowdsourcing experiments (see the fourth row and the third column in Figure~\ref{tsne}) which reflects the decision standard of the crowdsourcing workers. 

Table~\ref{tab3} indicates that our proposed model learned the settings of worker preferences in simulation experiments.  For example, worker 2 used the weight of color as $0.7$, and weight of number as $0.3$ in simulation setting 3. It may be observed that view 1 corresponded to number in simulation setting 3, and therefore the value $0.4281$ can be considered correct. 

\section{Conclusions}
In this study, we investigated a new end-to-end framework designed to learn multiview representation embeddings from crowdsourced triplets data.
Based on the hypothesis that different crowd workers may have different views and the same crowd worker may choose different views in different tasks, we adopted triplet entropy and worker models to give different views different weights.
We convened 160 crowd workers to conduct experiments using two datasets in total.
The results demonstrated that our proposed method performed better in terms of multiple evaluation metrics on both simulated worker experiments and human crowdsourcing experiments using two datasets.
In our experiments, we chose ResNet18 as a baseline for comparison with our proposed approach. However, in future research, other network structures should be selected as baselines and compare the performance of the baseline method with and without multiview method.

Moreover, we confirmed that our multiview embeddings focused on different attributes of objects separately on an MINST with color and learned the preference of workers in the simulation experiments, as shown in Figure~\ref{tsne}. However, further study needs to investigate the semantics meaning of multiview embeddings in datasets with ambiguous views, such as the dog dataset considered here. Better performance could be achieved on our experimental dataset by setting the number of views, a hyperparameter, to be greater than or equal to $2$. However, methods of setting number of views should also be investigated. 

Our proposed crowd multiview method could become a typical solution for many other tasks when human might have multiple different views. For example, a user prefers a movie from several presenting movies when other favorite movies of the user are known. That is, the new preferred movie might be more similar to the previous favorite movies than other presenting movies. Such movie preferences data are another kind of relative comparisons. It is possible that users have multiple views to consider which movies are similar. Our method could be adopted if modifying the definition of relative comparison and adding multiple branches to other neural network structures.


\newpage

\begin{acks}
This work was supported by JST, the establishment of university fellowships towards the creation of science technology innovation, Grant Number JPMJFS2123,
the Research Grant for Young Scholars funded by Yamanashi Prefecture,
and by JST CREST, Grant Number JPMJCR21D1.

We thank Mr. Chengyang Qian (Shanghai Jiao Tong University) for discussing with us.

\end{acks}

\bibliographystyle{ACM-Reference-Format}
\bibliography{reference}

\newpage

\appendix
{\onecolumn
\section{Ablation study of removing triplet entropy}
We assumed that crowd workers might choose different views in different tasks or a different times and tended to choose a view that simplifies their decision making process for a given task. To measure the difficulty of tasks, we added triplet entropy to the view weights, i.e., $q^v_{{ijk}_{(m)}}  = \widetilde{h}^v_{ijk}  + w_m^v$ in Eq.~(\ref{q}). To evaluate the effectiveness of our proposed triplet entropy, we found a decline in performance by removing the triplet entropy from the view weights, i.e., $q^v_{{ijk}_{(m)}}  =  w_m^v$, as shown in Table~\ref{abla}.

\begin{table*}[!h]

\caption{Results on real crowdsourcing experiments on the 10-color MINST and Dog test sets with $2$ views. A larger value indicates better performance.  }

\centering
\begin{tabular}{|cccccccc|}
\hline
\multicolumn{1}{|c|}{Triplet Entropy}            & \multicolumn{1}{c|}{Accuracy}        & \multicolumn{4}{c|}{Clustering Eval.}                                                                                                                     & \multicolumn{1}{c|}{Linear Eval.}    & $2$-anchors Eval. \\ \hline
\multicolumn{2}{|c|}{\multirow{2}{*}{}}                                       & \multicolumn{2}{c|}{K-means}                                                & \multicolumn{2}{c|}{Agglomerative}                                          & \multicolumn{2}{c|}{\multirow{2}{*}{}}                   \\ \cline{3-6}
\multicolumn{2}{|c|}{}                                                        & \multicolumn{1}{c|}{Purity}          & \multicolumn{1}{c|}{NMI}             & \multicolumn{1}{c|}{Purity}          & \multicolumn{1}{c|}{NMI}             & \multicolumn{2}{c|}{}                \\ \hline
\multicolumn{8}{|c|}{10-color MNIST}                                                                                                                                                                                                                                                                 \\ \hline
\multicolumn{1}{|c|}{\ding{55}}                & \multicolumn{1}{c|}{0.6133}          & \multicolumn{1}{c|}{0.4831}          & \multicolumn{1}{c|}{0.7157}          & \multicolumn{1}{c|}{0.4963}          & \multicolumn{1}{c|}{0.7382}          & \multicolumn{1}{c|}{0.6981}          & 0.5022               \\ \hline
\multicolumn{1}{|c|}{\ding{51}} & \multicolumn{1}{c|}{\textbf{0.6941}} & \multicolumn{1}{c|}{\textbf{0.5461}} & \multicolumn{1}{c|}{\textbf{0.7539}} & \multicolumn{1}{c|}{\textbf{0.5587}} & \multicolumn{1}{c|}{\textbf{0.7784}} & \multicolumn{1}{c|}{\textbf{0.8243}} & \textbf{0.5899}   \\ \hline
\multicolumn{8}{|c|}{Dog}                                                                                                                                                                                                                                                                            \\ \hline
\multicolumn{1}{|c|}{\ding{55}}                & \multicolumn{1}{c|}{{0.3942}} & \multicolumn{1}{c|}{{0.2462}} & \multicolumn{1}{c|}{{0.4323}} & \multicolumn{1}{c|}{{0.2677}} & \multicolumn{1}{c|}{{0.4691}} & \multicolumn{1}{c|}{{0.5211}} & {0.1772 }      \\ \hline

\multicolumn{1}{|c|}{\ding{51}} & \multicolumn{1}{c|}{\textbf{0.4882}} & \multicolumn{1}{c|}{\textbf{0.3122}} & \multicolumn{1}{c|}{\textbf{0.6402}} & \multicolumn{1}{c|}{\textbf{0.3351}} & \multicolumn{1}{c|}{\textbf{0.6544}} & \multicolumn{1}{c|}{\textbf{0.5444}} & \textbf{0.2941}   \\ \hline
\end{tabular}

\label{abla}
\end{table*}
}

\end{document}